\documentclass[final,5p,times,twocolumn]{elsarticle}

\usepackage{amssymb}

\usepackage{amsfonts,amssymb,amsmath,amsbsy}
\usepackage{physics}
\usepackage{color,calc}
\usepackage{bm}
\usepackage{array}
\usepackage{booktabs}
\usepackage{graphicx}
\usepackage{graphics}
\usepackage{eepic,epsfig}
\usepackage[breaklinks,hidelinks,colorlinks=true,linkcolor=blue,citecolor=blue,filecolor=blue,urlcolor=blue]{hyperref}
\usepackage{siunitx}
\usepackage{soul}

\usepackage[dvipsnames]{xcolor}
\definecolor{myblue}{RGB}{0, 0, 255}
\usepackage[toc,page]{appendix}

\usepackage{microtype}
\usepackage{multirow}
\usepackage{booktabs}
\usepackage{makecell}

\def\be{ \begin{equation} }
\def\ee{ \end{equation} }
\def\bea{ \begin{eqnarray} }
\def\eea{ \end{eqnarray} }
\def\bse{ \begin{subequations} }
\def\ese{ \end{subequations} }
\def\ba{ \begin{array} }
\def\ea{ \end{array} }
\def\bt{ \begin{tabular} }
\def\et{ \end{tabular} }

\def\to{\rightarrow}

\def\etal{\textit{et al.}}

\long\def\/*#1*/{}

\journal{Nuclear Instruments and Methods A}

\begin{document}

\begin{frontmatter}



\title{Partially Coherent X-Ray Oscilex Radiation from a FEL-Modulated Positron Bunch during Its Planar Channeling in a Crystalline Undulator}


\author[inst1,inst2]{Hayk L. Gevorgyan}
\cortext[cor1]{Corresponding author}
\ead{hayk.gevorgyan@aanl.am}
            
\affiliation[inst1]{organization={Experimental Physics Division, A.I. Alikhanyan National Science Laboratory (Yerevan Physics Institute)},
            addressline={2 Alikhanyan Brothers St.}, 
            city={Yerevan},
            postcode={0036}, 
            country={Armenia}}

\affiliation[inst2]{organization={Quantum Technologies Division, A.I. Alikhanyan National Science Laboratory (Yerevan Physics Institute)},
            addressline={2 Alikhanyan Brothers St.}, 
            city={Yerevan},
            postcode={0036}, 
            country={Armenia}}

\author[inst3]{Lekdar A. Gevorgian}

\affiliation[inst3]{organization={Matinyan Center for Theoretical Physics, A.I. Alikhanyan National Science Laboratory (Yerevan Physics Institute)},
            addressline={2 Alikhanyan Brothers St.}, 
            city={Yerevan},
            postcode={0036}, 
            country={Armenia}}

\begin{abstract}

The radiation emitted at zero angle by a microbunched positron bunch undergoing planar channeling in a crystalline undulator (CU) is studied. The bunch energy is assumed to be far above the threshold for radiation generation in the dispersive CU medium. Besides the usual ``hard'' undulator radiation produced by channeling oscillations (channeling undulator radiation) and by the CU bending (crystalline undulator radiation), a ``soft'' medium-polarization component also appears at zero angle due to the oscillations that excite atomic electrons. We refer to this soft component as \emph{Oscilex} (oscillationally-excited) radiation. Since the two types of oscillations have different frequencies, they yield two distinct frequency components of both undulator and Oscilex radiation. The Oscilex frequencies are set by the plasma frequency and the characteristic oscillation frequency and are, to high accuracy, independent of the positron energy. The CU period is chosen so that the radiation wavelength is not shorter than the microbunch length, ensuring coherent emission from microbunches and partially coherent Oscilex emission from the full bunch. Analytical expressions are obtained for the spectral line shapes and the number of photons of spontaneous Oscilex radiation. For partially coherent emission, Gaussian distributions are used for both the bunch and microbunches. Gain factors for the two Oscilex components, including longitudinal form-factors, and the total number of partially coherent photons are derived. A positron bunch with LCLS parameters, modulated by SASE XFEL, channeling between (1 1 0) planes of a periodically bent diamond crystal is analyzed. The number of spontaneously emitted Oscilex photons exceeds the number of positrons by $1\text{÷}2$ orders of magnitude, and the gain factors reach $10^3 \text{÷} 10^4$.

\end{abstract}




\end{frontmatter}


\section{Introduction}\label{sec:intro}
Monochromatic directed photon beams radiated by bunches of high-energy charged particles have a wide practical application. At the microwave frequency range, such photon beams can be obtained by oscillations of relativistic electrons of a bunch in a periodic electric field, proposed by Ginzburg \cite{ginzburg1947}. Developing this idea, Motz \cite{motz} proposed theory to study the radiation of relativistic electrons that arises when passing through an external periodic magnetic field. This phenomenon is realized and named by the Motz's group as undulator radiation \cite{motz1953}.

Madey has shown that the stimulated undulator radiation (a free electron laser (FEL)) enhancement is determined by the derivative of the spontaneous radiation line shape \cite{madey}. As a result of quantum recoil, the principle of detailed balance between the probabilities of stimulated emission and absorption is violated --- at a certain frequency, the probability of emission becomes greater than the probability of absorption. In the microwave range, this phenomenon was observed in Stanford experiment using a resonator (with mirrors) \cite{elias}. X-ray undulator radiation (XUR) was first considered in \cite{Korkhmazian1970}. The idea of FEL without a cavity was proposed by \cite{saldin} and later was named SASE XFEL.


The problem of undulator radiation formed in a dispersive medium was first theoretically studied in \cite{Gevorgyan1977}. It was shown that as the energy of the bunch approaches the threshold value for radiation formation, the radiation spectrum is generated at zero angle and narrows. This phenomenon was considered a discovery and is published in central journals \cite{Gevorgyan1979ab1,Gevorgyan1979ab2} after its registration as an invention \cite{Gevorgyan1979abpatent}. It can be concluded that the medium polarization similarly affects the channeling radiation \cite{BazylevZhevago1979}.

XUR in the presence of a dispersive medium was studied in \cite{Gevorgyan1977,Gevorgyan1979ab1,Gevorgyan1979ab2,Gevorgyan1979abpatent} and later investigated experimentally in \cite{Pantell1986}. The channeling radiation is considered in \cite{Kumakhov1976}. The theory of the crystalline undulator radiation (CUR), predicted in \cite{Kaplin1980}, has been developed without \cite{Korol1998} and with \cite{Avakian1998,Avakian2001} taking into account the medium polarization.

The coherent radiation is theoretically studied in \cite{Korkhmazian1977}. The asymmetry of a bunch leads to the radiation enhancement also in the short wavelength range. The effect of coherence of diffraction radiation at short wavelength was observed at the Japanese experiment \cite{Shibata1992}. By radiation spectrum, one can determine the real distribution of electrons in the bunch. The use of truncated electron bunch significantly increases FEL efficiency \cite{Gevorgyan1982} and the partial coherence is obtained for the cases of diffraction \cite{Mkrtchyan1998} and Smith-Purcell \cite{Saharian2001} radiations.

The research about the radiation problem of the bunch which density is modulated by laser beat waves (LBWs) was conducted in the paper \cite{Gevorgian1993}. The electron bunch, which interacts with LBWs, radiates coherently in the sub-millimeter range \cite{Shamamian2007}. In the paper \cite{Rosenzweig1995}, it was proposed to use partially coherent radiation at a resonant frequency to determine the modulation depth of the Linac Coherent Light Source (LCLS) electron bunch. Utilizing a stack of two plates as a microradiator immediately following SASE XFEL for determination of this parameter is theoretically considered in \cite{HGevorgyan2017}.

The results of article \cite{Gevorgian2013} show that one can use X-ray CUR for study of the microbunching process in XFEL and for the production of a monochromatic powerful photon beam. A review on the coherent X-ray radiation of various types produced by the modulated (microbunched) bunch in amorphous and crystalline radiators is given in \cite{Ispirian2013}.

When the bunch energy is much greater than the threshold energy for radiation generation in a dispersive medium of a CU, not only hard photons (CUR) are emitted, but also soft photons at zero angle. The frequency of these soft photons is independent of the bunch energy and is determined by the electron plasma frequency of the medium and the oscillation frequency of the charged particles within the bunch \cite{Gevorgian2005}. In the present article, due to its distinct properties, we coin the name \emph{oscillationally-excited} (shortly \emph{Oscilex}) \emph{radiation} to this distinct type of radiation. The abbreviation ``oscilex=oscillation+excitation'' captures the essence of radiation arising from oscillatory motion that induces excitation in a medium (a cold electron plasma).

The authors of \cite{Kostyuk2013} propose generating coherent hard CUR using a particle beam with a comb-like initial distribution, which is subsequently ``picobunched'' in a FEL. This effect is achievable with a superdense particle beam. In contrast, Oscilex radiation is significantly softer and can be more readily tuned to the FEL wavelength, enabling its generation using a particle beam with current densification and without the need for prior beam conditioning.

This paper is organized as follows. In Sec.~\ref{sec:theory} we present the general theory of undulator radiation. In Sec.~\ref{sec:CU}, we distinguish two types of radiation due to dispersive medium polarization --- CUR and Oscilex radiation, and derive formula for the photon number of two components of spontaneous Oscilex radiation from a single positron, one due to CU and another due to channeling oscillations. In Sec.~\ref{sec:coherent}, we present the theory of coherent radiation and the form-factors of a symmetric bunch with Gaussian longitudinal and transverse density distributions. In Sec.~\ref{DIR-mod}, we present partially coherent radiation gains of spontaneous Oscilex radiation components from a modulated bunch in Subsec.~\ref{DIR-mod-ch} due to (1) channeling and in Subsec.~\ref{DIR-mod-CU} due to (2) CU oscillations. In Sec.~\ref{sec:num}, we show our numerical results for this radiation in the case of positron bunch with LCLS parameters and modulated in SASE XFEL. Finally, Sec.~\ref{sec:concl} presents the conclusions.

\section{\label{sec:theory} Theory of Undulator Radiation}

The observed periodic \emph{sinusoidal}-like \cite{Bellucci2003} bending of the crystallographic planes enables us to find an analytic formula for the radiation generated in a CU using Jacobi–Anger expansion. Hereby, under the assumption of preserved channeling, an average trajectory and, hence, an average motion of a charged particle along the bent crystallographic planes (CU oscillations) can be considered as 
\bse\label{1}
\begin{align}
\vb*{r} (t) = \vb*{e}_y A \cos{\Omega t} + \vb*{e}_z \beta_{\parallel} c t, \label{1a} \\
\vb*{\beta} (t) = - \vb*{e}_y \beta_\perp \sin{\Omega t} + \vb*{e}_z \beta_{\parallel}, \label{1b}
\end{align}
\ese
where $\beta_\perp c= A \Omega$ is the maximum value of the transverse velocity of a charged particle, $\beta_\parallel c$ is an average longitudinal velocity of positrons, $\Omega = 2\pi \beta_\parallel c/l$ is an oscillation frequency, $l$ is a spatial period of CU, $A$ is a CU bending amplitude, $c$ is a velocity of light, $\vb*{e}_y$ and $\vb*{e}_z$ are the unit vectors along the transverse $y$ and longitudinal $z$ directions, respectively. There are two considerations in \eqref{1}: \begin{enumerate}
\item Radiation energy losses are negligible and are much less than the initial energy $\gamma m_e c^2$ of a charged particle, where $\gamma = 1/\sqrt{1-\beta^2}$ is a Lorentz factor, $\beta c$ is an initial velocity (before an undulator), $m_e$ is a rest mass of the same. 
\item Constancy of longitudinal velocity $\beta_\parallel = \sqrt{\expval{\beta^2_\parallel(t)}}$. This leads to the expression $\gamma_\parallel = 1/\sqrt{1 - \beta^2_\parallel} = \gamma/\sqrt{1+q^2/2}$ for the longitudinal Lorentz factor, where the following formulas are used: $\beta^2 = \expval{\beta^2_\parallel (t)} + \expval{\beta^2_\perp(t)}$ and $\expval{\beta^2_\perp(t)} = \expval{\beta^2_\perp \sin^2{\Omega t}} = \beta^2_\perp/2$. Thus, the longitudinal energy is less than the total energy by the factor of $\sqrt{1+q^2/2}$ times, where $q = \beta_\perp \gamma$ is called the undulator radiation parameter.
\end{enumerate}
The energy radiated per unit solid angle per unit frequency interval \cite{Jackson1999} is given by the following formula:
\bse\label{2} 
\begin{gather}
\frac{d^2 W}{d\omega dO} = \frac{e^2\sqrt{\varepsilon} \omega^2} {4\pi^2c} \left\lvert \vb*{A} (\omega) \right\rvert^2, \\
\vb*{A}(\omega) = \int\limits_{-T/2}^{T/2} \left(\vb*{n} \cp \vb*{\beta}(t)\right) e^{i(\omega t - \vb*{k} \cdot \vb*{r}(t))} dt,
\end{gather}
\ese
where $\vb*{A}(\omega)$ characterizes the radiation field, $\vb*{n}$ is a unit vector directed from the position of the charge to the observation point, representing direction of the instantaneous energy flux (Poynting's vector), $\vb*{k} = \sqrt{\varepsilon} \omega \vb*{n}/c$ is the wave vector of radiated photon with the frequency $\omega$ in the medium with dielectric permittivity $\varepsilon$, $e$ is an elementary charge, $T$ is the interaction time, and $dO = \sin{\vartheta} d\vartheta d\varphi$ represents the differential solid angle element in spherical coordinates, where $\vartheta$ and $\varphi$ are the polar and the azimuthal angles of radiation respectively.

Representation of $\vb*{n}$ in a spherical coordinates \eqref{3a} leads to the expression \eqref{3b}  
\bse\label{3} 
\begin{gather}
\vb*{n} = \vb*{e}_x \sin{\vartheta} \cos{\varphi} + \vb*{e}_y \sin{\vartheta} \sin{\varphi} + \vb*{e}_z \cos{\vartheta}, \label{3a} \\
\left(\vb*{n} \cp \vb*{\beta}(t)\right) = \vb*{a} - \vb*{b}  \sin{\Omega t}, \label{3b} \\
\vb*{a} = \beta_\parallel \left(\vb*{e}_x \sin{\vartheta} \sin{\varphi} - \vb*{e}_y \sin{\vartheta} \cos{\varphi}\right),  \notag\\
\vb*{b} = \beta_\perp \left( - \vb*{e}_x \cos{\vartheta} + \vb*{e}_z \sin{\vartheta} \cos{\varphi} \right), \notag
\end{gather}
\ese
where the formula \eqref{1b} is used. Another simplification is 
\be\label{4} 
\begin{gathered}
\omega t - \vb*{k} \vb*{r} = \alpha_1 t - \alpha_2 \cos{\Omega t}, \\
\alpha_1 = \omega \left(1 - \beta_\parallel \sqrt{\varepsilon} \cos{\vartheta} \right), \\
\alpha_2 = \frac{\omega}{\Omega} \beta_\perp \sqrt{\varepsilon} \sin{\vartheta} \sin{\varphi}, 
\end{gathered}
\ee
Hence, the radiation field during the time interval $T$ is given by the integral:
\be\label{5}
\vb*{A} = \int\limits_{-T/2}^{T/2} \left(\vb*{a} - \vb*{b} \sin{\Omega t}\right) e^{i(\alpha_1 t - \alpha_2 \cos{\Omega t})} dt .
\ee
As usual \cite{Gevorgyan1977,Gevorgyan1979ab1,Gevorgyan1979ab2,Gevorgyan1979abpatent}, one uses the expansion of a plane wave in (cylindrical) Bessel functions of the first kind, so called Jacobi-Anger expansion 
\be\label{6}
e^{-i \alpha_2 \cos{\Omega t}} = \sum^{\infty}_{m=-\infty} (-i)^m J_m (\alpha_2) e^{-i m \Omega t},
\ee
for integration of \eqref{5}. Using Euler's formula and \eqref{6}, one gets three terms in the expansion
\begin{multline}\label{7}
\vb*{A} = \sum_{m=-\infty}^\infty (-i)^m \left( \vb*{a} J_m(\alpha_2) + \frac{\vb*{b}}{2} \left(J_{m+1} (\alpha_2) + J_{m-1} (\alpha_2)\right) \right) \cdot \\
\cdot \int_{-T/2}^{T/2} e^{i(\alpha_1 - m \Omega) t} dt ,
\end{multline}
which can be simplified using a recurrence formula for Bessel functions of the first kind
\be\label{8}
J_{m+1} (\alpha_2) + J_{m-1} (\alpha_2) = \frac{2m}{\alpha_2} J_m(\alpha_2).
\ee
Considering $n$ oscillations in the length of a crystal $L = n l$, where $n$ is called the CU period number, a particle passes the CU in time $T = 2\pi n/\Omega$. Thus, in accordance to \eqref{8}, formula \eqref{7} becomes
\bse\label{9} 
\begin{gather}
\vb*{A} = \sum_{m=-\infty}^{\infty} \vb*{A}_m, \label{9a} \\
\vb*{A}_m = \frac{2\pi^2}{\Omega} (-i)^m J_m(\alpha_2) \frac{\sin{n x_m}}{\pi x_m} \left(\vb*{a} + \frac{m \vb*{b}}{\alpha_2}\right), 
\label{9b} \\
x_m = \pi \frac{\omega}{\Omega} \left(1 - \beta_\parallel \sqrt{\varepsilon} \cos{\vartheta} - \frac{m \Omega}{\omega} \right) . \label{9c}
\end{gather}
\ese

Radiation must occur also for the infinitely large number of oscillations $n\gg 1$ in a bent crystal, therefore, there is a natural condition to consider only the radiation field harmonics $\vb*{A}_m$ for which $x_m = 0$, since they in major contribute to the radiation. This condition ($x_m = 0$), accurate up to small energy losses due to radiation, corresponds to the law of energy-momentum conservation; therefore, $m\geq 0$. Under the condition $\beta_\parallel \sqrt{\epsilon} \geq 1$, the harmonic with index $m=0$ defines Vavilov-\v{C}erenkov radiation.

In \eqref{2} in the sum $ \left\lvert \vb*{A} \right\rvert^2 = \sum\limits_{m,m^\prime=1}^{\infty} \vb*{A}_m \vb*{A}^\ast_{m^\prime}$ only equiharmonic terms remain, since $x_m$ and $x_{m^\prime}$ are not simultaneously equal to zero for mixed (non-equiharmonic) terms. Thus, formula \eqref{2} becomes

\begin{multline}\label{10} 
\frac{d^2 W}{d\omega dO} = \frac{e^2\sqrt{\varepsilon} \omega^2} {c\Omega^2} \sum_{m=1}^{\infty} \left(\beta_\parallel^2 \sin^2 {\vartheta} - \frac{m}{\alpha_2} \beta_\parallel \beta_\perp \sin{2\vartheta} \sin{\varphi} + \right. \\
\left. + \frac{m^2}{\alpha_2^2} \beta_\perp^2 \left(1-\sin^2{\vartheta} \sin^2{\varphi}\right) \right) J_m^2(\alpha_2) \frac{\sin^2{(n x_m)}}{x_m^2}.
\end{multline}

\section{Oscilex Radiation}\label{sec:CU}

The relativistic positrons, falling parallel to the planes of crystal, interact with many atoms of crystal simultaneously. Therefore, between the crystal planes they move in the average potential field of these atoms. The harmonic potential is a good approximation for this field \cite{Bazilev1987}. The positrons are oscillating according to the harmonic law with the frequency depending on their energy. This phenomenon of channeling is conserved at the falling angles smaller than the Lindhard critical angle $\Theta_{L}=\sqrt{2\nu/\gamma}$, where $\nu$ and $\gamma$ are the potential barrier height $U_0$ and the positron energy $E$ in units of the rest energy of a positron $mc^2$, respectively.\par
The channeling phenomenon will take place for the all falling positrons of a bunch, if the maximum angle of the crystal bending $\beta_\perp$ is smaller than the Lindhard angle $\Theta_{L}$ (the \emph{channeling conservation condition} in CU)
\be
\frac{2\pi A}{l} < \sqrt{2\nu/\gamma}.
\ee



The dielectric permittivity in the dispersive medium of CU is given by the following formula
\be\label{11} 
\varepsilon (\omega) = 1 - \left(\frac{\omega_p}{\omega}\right)^2, \quad \omega \gg \omega_p,
\ee
where $\omega_p$ is a plasma frequency of a CU crystal. Note that the medium polarization, represented by formula~\eqref{11}, is responsible for the emergence of the Oscilex radiation, which has no analog in a conventional vacuum-filled undulator.

Since we are interested in the characteristics of the fundamental harmonic of Oscilex radiation at zero angle, generated as a result of the interaction of a relativistic positron, oscillating with a frequency $\Omega$ of CU, with the atomic electrons of the medium, we use the following expansions ($\vartheta \ll 1$)

\be
\beta_\parallel \approx 1 - 1/\left(2\gamma_\parallel^2\right), \quad \sqrt{\varepsilon} \approx 1 - \omega_p^2/\left(2\omega^2\right), \quad \cos{\vartheta} \approx 1 - \vartheta^2/2.
\ee
For the fundamental harmonic $x_1 \overset{\Delta}{=} x (\omega, \theta)$, where $\theta = \vartheta \gamma_\parallel$, one gets
\be
x(\omega, \theta) = \frac{\pi}{\omega \omega_{max}} \left(\left(1+\theta^2\right) \omega^2 - \omega_{max} \omega + \left(\omega_p \gamma_\parallel\right)^2 \right), \quad \omega_{max} = 2\Omega \gamma^2_\parallel.
\ee
The frequencies corresponding to the peak values of the radiation line shapes, determined by the equation $x(\omega, \theta) = 0$, are given by
\be
\frac{\omega_{max}}{2(1+\theta^2)} \left(1 \mp \sqrt{1 - \frac{\gamma^2_{th} (\theta)}{\gamma^2_\parallel}}\right), \quad \gamma_{th} (\theta) = \frac{\omega_p}{\Omega} \sqrt{1 + \theta^2}.
\ee
For energies much greater than the threshold energy for radiation generation $\gamma_\parallel \gg \gamma_{th} (\theta)$, we distinguish two types of radiation at zero angle due to medium polarization \cite{Gevorgyan1977,Gevorgyan1979ab1,Gevorgyan1979ab2,Gevorgyan1979abpatent,Gevorgian2013,Gevorgian2005,avakian2013,HGevorgyan2021a, HGevorgyan2021b, HGevorgyan2015, HGevorgyan2024a}: hard CUR at frequency $\omega_{+}(\theta)$, which, as in a conventional vacuum-filled undulator, depends on the energy of ultrarelativistic charged particles $\gamma$; and Oscilex radiation at frequency $\omega_{-}$, which primarily depends on the plasma frequency $\omega_p$, a property of the dispersive medium, on the characteristic oscillation frequency $\Omega$, and with high precision is independent of $\gamma$. The frequencies of Oscilex and undulator radiations are respectively:
\be
\omega_- = \frac{\omega^2_p}{2\Omega}, \quad \omega_+ (\theta) = \frac{\omega_{max}}{1+\theta^2},
\ee
where $\omega_{max}$ is the maximum frequency of undulator radiation emitted at zero angle. 

In a CU, undulator radiation is generated both due to the oscillations of particles during channeling and at the frequency of the CU itself. Naturally, when determining $\gamma_\parallel$ in terms of $\gamma$, the influence of both oscillations must be taken into account. However, this work is dedicated to the study of Oscilex radiation, which has a frequency independent of $\gamma$.


For the frequency-angular distribution of the number of Oscilex radiation photons of the fundamental harmonic $\tfrac{d^3 N_{ph}}{d\omega dO} = \tfrac{d^3 W}{d\omega dO} \tfrac{1}{\hbar \omega}$, to the order of small terms of $\vartheta^2$, we have:
\be
\frac{d^3 N_{ph}}{d\omega dO} = \frac{\alpha \omega}{4\Omega^2} \beta^2_\perp \left(\cos^2{\varphi} + \left(1 - \vartheta^2 \frac{\omega}{\Omega}\right)^2 \sin^2{\varphi} \right) \frac{\sin^2{(n x(\omega, \theta))}}{x^2 (\omega, \theta)},
\ee
where $J_1 (\alpha_2) \approx \alpha_2 /2$ is used as $\vartheta \to 0$, $\hbar = h/(2\pi)$ is the reduced Planck constant, and $\alpha = e^2 / (\hbar c) = 1/137$ is a fine structure constant.

For the frequency distribution of Oscilex radiation photons at an angle $\vartheta = 0$, given that radiation at zero angle in the forward direction would be part of the upper hemisphere $\int \, dO = \int_{0}^{2\pi} \int_{0}^{\pi/2} \sin{\vartheta} \, d\vartheta \, d\varphi = 2\pi$, one gets:
\be
\begin{gathered}
\frac{dN_{ph}}{d\omega} = 2\pi\alpha \left(\frac{\beta_\perp}{2\Omega}\right)^2 \omega \phi(x), \quad \phi(x) = \frac{\sin^2{(n x)}}{x^2}, \\ 
x \overset{\Delta}{=} x(\omega, 0) = \frac{\pi}{\omega \omega_{max}} (\omega - \omega_-) (\omega_{max} -\omega).
\end{gathered}
\ee
For $n\gg 1$, with accuracy up to small terms of order $1/n$, the representation $\lim\limits_{n\to \infty} {\sin^2{(n x)}/x^2} = \pi n \delta(x)$ can be used. The area under the curve of $\phi(x)$ is approximated by an isosceles triangle with height $\phi(0) = n^2$ and base $2 \pi/n$ (where $\phi (\pm \pi/n) = 0$), where $\delta(x)$ denotes the Dirac delta function. The relative linewidth (defined as the spectral half-width at zero level normalized to the central frequency) of the line shape \cite{HGevorgyan2024a} function $\sin^2{(n x)}/x^2$ is equal to $\Delta\omega/\omega_{-}=\Delta\omega/\omega_{m}=1/n$.

For $\omega \approx \omega_-$, and considering $\omega_- \ll \omega_{max}$, we have $x = \pi/n$ ($ (\omega -\omega_-) / \omega = 1/n$). Therefore, to the order of small terms $1/n$, we have ($\phi(0) = n^2$, $n \gg 1$):

\be
\phi(x) = \pi n \delta{\left(\pi \frac{\omega - \omega_-}{\omega}\right)} = n \omega \delta{(\omega-\omega_-)}.
\ee
Thus, the number of spontaneously emitted Oscilex photons generated by a single positron undergoing $n$ oscillations with frequency $\Omega$ in a dispersive medium follows:
\be\label{N_sp}
N_{sp} = 2\pi n \alpha \beta^2_\perp \left(\frac{\omega_-}{2\Omega}\right)^2 = 2\pi n \alpha \beta^2_\perp \left(\frac{\omega_p}{2\Omega}\right)^4 = 2\pi n \alpha \beta^2_\perp \left(\frac{l}{2\lambda_p}\right)^4.
\ee
Besides the Oscilex photons with a wavelength $\lambda_- = 2 \lambda^2_p / l$ ($\omega_- = \omega^2_p/2\Omega$) resulting from the $n$ oscillations of positrons at the CU frequency $\Omega$, photons of the same type are also radiated at zero angle with a wavelength $\lambda_{ch} = 2 \lambda^2_p / l_{ch}$ ($\omega_{ch} = \omega^2_p/2\Omega_{ch}$) due to oscillations at the channeling frequency $\Omega_{ch}$. In this case, the replacements $\beta_\perp \to \beta_\perp (s)$, $\omega_- \to \omega_{ch}$, $\Omega \to \Omega_{ch}$, $l \to l_{ch}$, and $n \to n_{ch}$ are required in Eq.~\eqref{N_sp}.

\section{Partially Coherent Radiation and Form-Factors of a Bunch}\label{sec:coherent}

The formula for frequency-angular average distribution of the number of photons of any nature, radiated by freeform positron bunch, has the following form \cite{Korkhmazian1977}
\be\label{16}
\begin{gathered}
N_{tot}(\omega,\vartheta)=N_{incoh}(\omega,\vartheta)+N_{coh}(\omega,\vartheta),\\
N_{coh}(\omega,\vartheta)=N^2_{b} F N_{ph}(\omega,\vartheta),
\end{gathered}
\ee
where $N_{ph}(\omega,\vartheta)$ represents the frequency-angular distribution of the number of photons radiated by a single positron, and $N_{b}$ denotes the number of positrons in a bunch. A microbunch radiates coherently, when $F \approxeq 1$ (the \emph{coherence condition} of radiation). \par
The bunch's form-factor $F=F_{Z}(\omega,\vartheta)\times F_{R}(\omega,\vartheta)$ is determined by longitudinal $F_{Z}(\omega,\vartheta)$ and transverse $F_{R}(\omega,\vartheta)$ form-factors
\be\label{17}
\begin{gathered}
F_{Z}(\omega,\vartheta)= \left\lvert h^2_Z \right\rvert, \quad h_Z \overset{\Delta}{=} \int\limits_{-\infty}^{\infty} f(Z) e^{-i\frac{\omega}{c} Z \cos{\vartheta}} dZ ,\\
F_{R}(\omega,\vartheta)=  \left\lvert h^2_R \right\rvert, \quad h_R \overset{\Delta}{=} \int\limits_{-\infty}^{\infty} f(R) e^{-i\frac{\omega}{c} R \sin{\vartheta}} dR ,
\end{gathered}
\ee
where $f(Z)$ and $f(R)$ are the positron bunch's density distributions in longitudinal $Z$ and transverse $R$ directions, respectively.\par







When $f(Z)$ and $f(R)$ are Gaussian distribution functions with variances $\sigma^2_{Z}$ and $\sigma^2_{R}$: $1 / \left(\sqrt{2\pi} \sigma_Z\right) \exp{- Z^2 / \left(2 \sigma^2_Z\right)}$ and $1 / \left(\sqrt{2\pi} \sigma_R\right) \exp{- R^2 / \left(2 \sigma^2_R\right)}$, respectively, we have
\be\label{18}
\begin{gathered}
F_{Z}(\omega,\vartheta)=\exp{-\left(\frac{\omega \sigma_Z \cos{\vartheta}}{c}\right)^2}, \\
F_{R}(\omega,\vartheta)=\exp{-\left(\frac{\omega \sigma_R \sin{\vartheta}}{c}\right)^2}.
\end{gathered}
\ee
Therefore form-factors at zero angle are the following
\begin{equation}\label{19}
F_{Z}(\omega)=\exp{-\left(\frac{\omega \sigma_Z}{c}\right)^2}, \qquad F_{R}=1.
\end{equation}
Later, we will use the notation $\sigma \overset{\Delta}{=} \sigma_Z$.

\section{Partially Coherent X-ray Radiation from a Microbunched Bunch in a CU}\label{DIR-mod}
Let a bunch with $N_b$ particles have a Gaussian distribution with a standard deviation $\sigma$. As a result of particle grouping, the bunch transforms into $2\sigma/\lambda_r$ microbunches, each with a Gaussian distribution of standard deviation $\lambda_r/2$. The microbunches, containing $N_b / (2\sigma/\lambda_r)$ particles, will radiate coherently at a wavelength $\lambda$ with a form-factor (a coherence factor) \cite{Korkhmazian1977}
\be 
F(\lambda) = \exp{- \left(\frac{\pi}{2} \frac{\lambda_r}{\lambda}\right)^2}.
\ee
The number of partially coherent radiated photons results in the expression
\be
N_{coh} = \frac{2\sigma}{\lambda_r} N^2_b \left(\frac{\lambda_r}{2\sigma}\right)^2 F(\lambda) N_{sp} = K(\lambda) N^b_{sp}, \quad K(\lambda) = \frac{\lambda_r}{2\sigma} N_b F(\lambda),
\ee
where $N^b_{sp} = N_b N_{sp}$ is the number of spontaneously radiated photons by the bunch, and $K(\lambda)$ is the gain factor of spontaneous radiation. 

\subsection{Partially Coherent Oscilex Radiation due to Channeling Oscillations}\label{DIR-mod-ch}

The number of channeling oscillations is given by $n_{ch} = L_D / l_{ch}$, $\beta_\perp (s) = s \Theta_L$, where $s$ ($0 \leq s \leq 1$) is the amplitude of the channeled positron oscillations in units of $d/2$, half of the channel width. The average squared transverse velocity is given by $\langle \beta^2_\perp (s) \rangle = \int_{0}^{1} \beta^2_\perp (s) \, ds =\Theta^2_L/3$, and the channeling radiation wavelength is expressed as $\lambda_{ch} = 2 \lambda^2_p / l_{ch}$. The number of photons radiated by a channeled positron oscillating with amplitude $s$ follows
\be
N_{sp} (ch) = \frac{\pi \alpha n_{ch}}{8} \Theta^2_L \left(\frac{l_{ch}}{\lambda_p}\right)^4 s^2.
\ee
The number of spontaneously radiated photons in Oscilex radiation due to oscillations at the channeling frequency, with wavelength $\lambda = \lambda_{ch}$, results in the expression
\be
N^b_{sp} (ch) = \frac{\pi \alpha n_{ch}}{24} \Theta^2_L \left(\frac{l_{ch}}{\lambda_p}\right)^4 N_b. 
\ee
The number of partially coherent radiated photons and the corresponding relative spectral linewidth are expressed as
\be
N_{coh} (ch) = N^b_{sp} (ch) K(\lambda_{ch}) = N^b_{sp} (ch) \frac{\lambda_r}{2\sigma} N_b F(\lambda_{ch}), \quad \frac{\Delta \lambda_{ch}}{\lambda_{ch}} = \frac{1}{n_{ch}}
\ee
where the form-factor is given by $F(\lambda_{ch}) = \exp{- (\tfrac{\pi}{2} \tfrac{\lambda_r}{\lambda_{ch}})^2 }$.

\subsection{Partially Coherent Oscilex Radiation due to CU Oscillations}\label{DIR-mod-CU}

The number of spontaneously radiated photons in Oscilex radiation due to oscillations at the undulator frequency, with wavelength $\lambda = \lambda_r$, results in the expression
\be
N^b_{sp} = \frac{\pi \alpha n}{8} \beta^2_\perp \left(\frac{l}{\lambda_p}\right)^4 N_b,
\ee
where $n = L_D/l$, $L_D$ is the dechanneling length, $l$ is the spatial period of the undulator, $\beta_\perp = 2 \pi A /l$, and $A$ is the oscillation amplitude. The number of partially coherent radiated photons and the corresponding relative spectral linewidth are expressed as
\be
N_{coh} = N^b_{sp} K(\lambda_r) = N^b_{sp} \frac{\lambda_r}{2\sigma} N_b F(\lambda_r), \quad \frac{\Delta \lambda}{\lambda} = \frac{1}{n},
\ee
where the form-factor is equal to $F(\lambda_r) = \exp{-\tfrac{\pi^2}{4}} = 0.084805$.

\section{Numerical Results}\label{sec:num}

Numerical results for the case when the modulated positron bunch is incident perpendicular to the CU-section (see Fig.~\ref{fig:CU}) are shown below. The numerical calculations are performed for a diamond and the positron bunch with the LCLS parameters and modulated in the SASE XFEL process.\par
The modulated positron bunch parameters are: $N_{b}=1.56\times 10^{9}$, $E=13.6$~\SI{}{\giga\electronvolt} $(\gamma=2.66\times 10^{4})$, $\hbar\omega_{r}=8.3$~\SI{}{\kilo\electronvolt}
($\lambda_{r}=1.5$~\SI{}{\angstrom}), $\sigma_{Z}=9\times 10^{-4}$~\SI{}{\centi\meter}. \par 
Positrons are channeled between the (1 1 0) crystallographic planes of a diamond crystal, with an interplanar distance $d=1.25865$~\SI{}{\angstrom}. Then the potential barrier height is $U_{0}=24.9$~\SI{}{\electronvolt} ($\nu=U_{0}/m c^{2}=4.87\times 10^{-5}$) \cite{Bazilev1987}, and the energy of plasma oscillations in a diamond medium is $\hbar\omega_{p}=38$~\SI{}{\electronvolt} ($\lambda_{p}=3.2763\times 10^{-6}$~\SI{}{\centi\meter}) \cite{Gevorgian2007}.

CU represents the periodically bent monocrystal of a diamond with the spatial period $l=1.43\times 10^{-3}$~\SI{}{\centi\meter} ($\lambda_{-}=\lambda_{r}=2\lambda^2_{p}/l=1.5$~\SI{}{\angstrom}), the amplitude $A=1$~\SI{}{\angstrom} and $\beta_\perp = 2\pi A/l = 4.39\times 10^{-5}$. Note that the components of Oscilex radiation are generated due to the oscillations of positrons (1) at the channeling frequency and (2) at the CU frequency. The number of oscillations is limited by dechanneling length $L_{D} = 0.816$~\SI{}{\centi\meter} \cite{Gevorgian2013, Biryukov2007}, which is considered equal to demodulation length \cite{Kostyuk2011}. The possibilities and methods to obtain sinusoidal bending of CU are discussed in \cite{Kostyuk2010}.

\ul{The parameters of channeled positrons and their \emph{channeling conservation condition} in CU}: the amplitudes of channeled positrons are given by $A(s) = (d/2) s$, where $0\leq s \leq 1$, the spatial period is $l_{ch} = l_0 \sqrt{\gamma}$, $l_0 = \pi d/\sqrt{2\nu} = 4\times 10^{-6}$~\SI{}{\centi\meter} ($\sqrt{2\nu} = 9.87 \times 10^{-3}$). The Lindhard channeling angle is $\Theta_{L}=\sqrt{2\nu/\gamma} = 6.05 \times 10^{-5}$ ($\sqrt{\gamma} = 1.63 \times 10^2$). Consequently, the spatial period is $l_{ch} = 6.53 \times 10^{-4}\SI{}{\centi\meter}$, and the wavelength is $\lambda_{ch} = 3.285$~\SI{}{\angstrom}.

As a result, one gets the numerical results for radiation characteristics of partially coherent Oscilex radiation due to (1) channeling and (2) CU oscillations, illustrated in Table~\ref{Table:NumResults}.  

\begin{table*}[tbph]
\centering
\begin{tabular}{|c|c|c|c|c|c|c|c|}
\hline
\makecell{\textbf{Oscilex}\\\textbf{radiation}\\\textbf{component}} & 
\makecell{\textbf{Number of}\\\textbf{oscillations}} & 
\multicolumn{3}{c|}{\makecell{\textbf{Number of photons}}} & 
\makecell{\textbf{Coherence}\\\textbf{factor}} & 
\makecell{\textbf{Gain}\\\textbf{factor}} & \makecell{\textbf{Relative}\\\textbf{linewidth}} \\
\cline{3-5}
 & & 
\multicolumn{2}{c|}{\makecell{\textbf{Spontaneous}\\\textbf{radiation}}} & 
\makecell{\textbf{Coherent}\\\textbf{radiation}} & & & \\
\cline{3-4} 
 & & \textbf{Single positron} & \textbf{Positron bunch} & & & & \\
\hline
(1) Channeling & \makecell{$n_{ch} = L_{D}/l_{ch}$\\$\approx 1248$} & \makecell{$N_{sp}(ch)$\\$\approx 20$} & \makecell{$N^b_{sp} (ch)$\\$= 1.08\times 10^{10}$} & \makecell{$N_{coh} (ch)$\\$= 8.38 \times 10^{13}$} & \makecell{$F(\lambda_{ch})$\\$= 0.597888$} & \makecell{$K_{ch}$\\$= 7772.55$} & \makecell{$\tfrac{\Delta \lambda_{ch}}{\lambda_{ch}}$\\$\approx 8 \times 10^{-4}$} \\
\hline
(2) CU & \makecell{$n=L_{D}/l$\\$\approx 570$} & \makecell{$N_{sp}$\\$\approx 114$} & \makecell{$N^b_{sp}$\\$= 1.79 \times 10^{11}$} & \makecell{$N_{coh}$\\$= 1.97 \times 10^{14}$} & \makecell{$F(\lambda_r)$\\$= 0.084805$} & \makecell{$K$\\$= 1102.46$} & \makecell{$\tfrac{\Delta \lambda}{\lambda}$\\$\approx 1.754 \times 10^{-3}$} \\
\hline
\end{tabular}
\caption{Radiation characteristics of partially coherent Oscilex radiation from a positron bunch with LCLS parameters, modulated in an SASE XFEL process and channeled along the (1 0 0) crystallographic planes of periodically bent monocrystal of a diamond.}
\label{Table:NumResults}
\end{table*}


\begin{figure}
  \includegraphics[width=\linewidth]{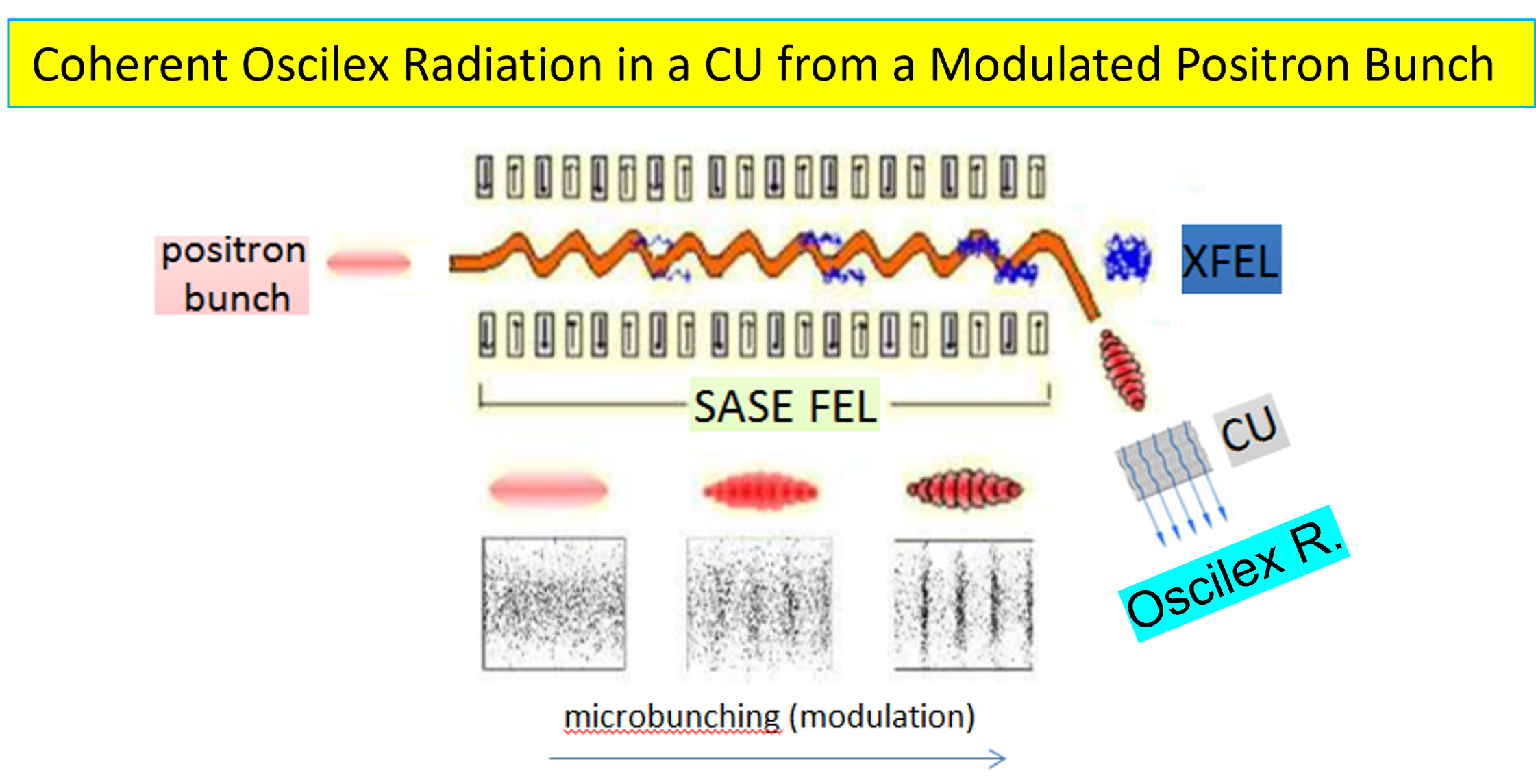}
  \caption{Scheme of the experiment to obtain a powerful, super-monochromatic, and directed X-ray photon beam. A schematic diagram of a single-pass Free Electron Laser (FEL) operating in the Self-Amplified Spontaneous Emission (SASE) mode is used [\href{https://arxiv.org/pdf/hep-ph/0106314}{TESLA Design Technical Report, March 2001, Part I, p. I-37}]  with the addition of a Crystalline Undulator (CU) and the Oscilex radiation generated within it. In the original figure, the bunch density modulation (microbunching), which develops in parallel with the SASE XFEL radiation power, was shown in the lower part.}
  \label{fig:CU}
\end{figure}

\section{Conclusion}\label{sec:concl}

It is shown that at energies exceeding the threshold energy for radiation in the dispersive medium of a crystalline undulator, a spontaneous radiation line shape is formed at zero angle in the soft region (Oscilex radiation), the frequency of which depends only on the plasma frequency of the medium and the oscillation frequency. Two components of Oscilex radiation are distinguished: those arising from (1) channeling oscillations and (2) CU oscillations. The wavelength of the (1) component is longer than that of the (2) component; therefore, the coherence factor of radiation from a modulated bunch is higher in case (1) than in case (2), while the number of spontaneously radiated photons is lower. The characteristics of partially coherent Oscilex radiation from a positron bunch with LCLS parameters, modulated in an SASE XFEL process and channeled along the (1, 0, 0) crystallographic planes of periodically bent monocrystal of a diamond, are investigated. It is shown that, due to partial coherence of the bunch, the number of Oscilex photons from positron bunch oscillations increases by three orders of magnitude for CU oscillations and by four orders for channeling oscillations (see Table~\ref{Table:NumResults}).

Since the intensity of the directed beams of Oscilex photons at different energies exceeds that of spontaneous radiation by several orders of magnitude, these powerful, directed, monochromatic beams of Oscilex photons may have significant practical applications and be utilized in scientific research \cite{Marangos2020, Boutet2018, Quiney2010, Krupyanskii2014, Huang2021, Vartanyants2015, Yabashi2013, Pellegrini2016}.

Notably, the considered phenomenon --- partially coherent Oscilex radiation from a microbunched bunch --- also takes place during axial channeling of electrons.

 \bibliographystyle{elsarticle-num} 

\begin{thebibliography}{00}


\bibitem{ginzburg1947} V. L. Ginzburg, ``Ob izluchenii mikroradiovoln i ikh pogloshchenii v vozdukhe (On the emission of microwaves and their absorption in air),'' Izvestiya Akad. Nauk SSSR, Ser. Fiz. \textbf{11}, 165 (1947).

\bibitem{motz} H. Motz, ``Applications of the radiation from fast electron beams,'' \href{https://doi.org/10.1063/1.1700002}{J. Appl. Phys. \textbf{22}, 527 (1951)}.


\bibitem{motz1953} H. Motz, W. Thon, and R. N. Whitehurst, ``Experiments on radiation by fast electron beams,'' \href{https://doi.org/10.1063/1.1721389}{J. Appl. Phys. \textbf{24}, 826 (1953)}.

\bibitem{madey}
J. M. J. Madey, ``Stimulated emission of bremsstrahlung in a periodic magnetic field,'' \href{https://doi.org/10.1063/1.1660466}{J. Appl. Phys. \textbf{42}, 1906 (1971)}.
  
\bibitem{elias}
L. R. Elias, W. M. Fairbank, J. M. Madey, H. A. Schwettman, and T. I. Smith, ``Observation of stimulated emission of radiation by relativistic electrons in a spatially periodic transverse magnetic field,'' \href{https://doi.org/10.1103/PhysRevLett.36.717}{Phys. Rev. Lett. \textbf{36}, 717 (1976)}.

\bibitem{Korkhmazian1970} 
N.~A.~Korkhmazian, ``Izluchenie bystrykh zaryazhennykh chastits v poperechnykh ėlektrostaticheskikh sinusoidal'nykh polyakh (Radiation of fast charged particles in transverse electric sinusoidal fields),'' \href{https://arar.sci.am/dlibra/publication/144362/edition/131352/content}{Izvestiya ASSR NA, Fizika \textbf{5}, 287 (1970)}; ``Generatsiya zhestkikh kvantov v ėlektrostaticheskikh ondulyatorakh (Radiation of fast charged particles in transversal electrostatic sinusoidal fields),'' \href{https://arar.sci.am/dlibra/publication/144385/edition/131373/content}{Izvestiya ASSR NA, Fizika \textbf{5}, 418 (1970)}.

\bibitem{saldin}
A. M. Kondratenko and E. L. Saldin, ``Generation of coherent radiation by a relativistic electron beam in an ondulator,'' \href{https://cds.cern.ch/record/1107977/files/p207.pdf}{Part. Accel. \textbf{10}, 207 (1980)}.





\bibitem{Gevorgyan1977}
L.~A.~Gevorgyan and N.~A.~Korkhmazyan, 
``Hard undulator radiation in a dispersive medium in the dipole approximation,'' YerPhI Scientific Reports \textbf{273}(6)--77 (1977).

\bibitem{Gevorgyan1979ab1}
L.~A.~Gevorgyan and N.~A.~Korkhmazyan, 
``Hard undulator radiation in a dispersive medium in the dipole approximation,'', \href{https://doi.org/10.1016/0375-9601(79)90251-2}{Phys. Lett. A \textbf{74}, 453 (1979)}.

\bibitem{Gevorgyan1979ab2}
L.~A.~Gevorgyan and N.~A.~Korkhmazyan, ``Undulatory radiation in dispersive media,'' Zh. Eksp. Teor. Fiz. \textbf{76}, 1226 (1979) [\href{http://jetp.ras.ru/cgi-bin/dn/e_049_04_0622.pdf}{J. Exper. Theoret. Phys. \textbf{49}, 622 (1979)}].

\bibitem{Gevorgyan1979abpatent}
L.~A.~Gevorgyan and N.~A.~Korkhmazyan, ``Sposob generatsii kvazimonokhromaticheskogo zhestkogo izlucheniya'' (``Method for generation of quasi-monochromatic hard radiation''), \emph{USSR Patent} \textnumero 784729 (1980).

\bibitem{BazylevZhevago1979}
V.~A.~Bazylev and N.~K.~Zhevago, ``Influence of the polarisation of the medium on the radiation of channeled particles,'' \href{https://doi.org/10.1016/0370-2693(79)90279-x
}{Phys. Lett. B \textbf{84}, 182 (1979)}. 

\bibitem{Pantell1986} 
R.~H.~Pantell, J.~Feinstein, A.~S.~Fisher, T.~L.~Deloney, M.~B.~Reid, and W.~M.~Grossman, ``Benefits and costs of the gas-loaded, free electron laser,'' \href{https://doi.org/10.1016/0168-9002(86)90901-0}{Nucl. Instr. and Meth. A \textbf{250}, 312 (1986)}. 

\bibitem{Kumakhov1976}
M.~A.~Kumakhov, ``On the theory of electromagnetic radiation of charged particles in a crystal,'' \href{https://doi.org/10.1016/0375-9601(76)90438-2}{Phys. Lett. A, \textbf{57}, 17 (1976)}.  

\bibitem{Kaplin1980} 
V.~V.~Kaplin, S.~V.~Plotnikov, and S.~A.~Vorob'ev, ``Charged particle radiation while channeling in deformed crystals,'' Zh. Eksp. Teor. Fiz. \textbf{50}, 1079 (1980).

\bibitem{Korol1998} 
A.~V.~Korol, A.~V.~Solov’yov, and W.~Greiner, ``Coherent
radiation of an ultra-relativistic charged particle channeled in
a periodically bent crystal,'' \href{https://doi.org/10.1088/0954-3899/24/5/001}{J. Phys. G \textbf{24}, L45 (1998)}.

\bibitem{Avakian1998} 
R.~O.~Avakian, L.~A.~Gevorgian, K.~A.~Ispirian, and R.~K.~Ispirian, Pisma Zh. Eksp. Teor. Fiz. \textbf{68}, 437 (1998).

\bibitem{Avakian2001} 
R.~O.~Avakian, L.~A.~Gevorgian, K.~A.~Ispirian, and R.~K.~Ispirian, ``Spontaneous and stimulated radiation of particles in crystalline and nanotube undulators,'' \href{https://doi.org/10.1016/S0168-583X(00)00181-6}{Nucl. Instr. and Meth. B \textbf{173}, 112 (2001)}.


\bibitem{Korkhmazian1977} 
N.~A.~Korkhmazian, L.~A.~Gevorgian, and M.~P.~Petrosian, ``Vliyanie plotnosti raspredeleniya ėlektronov na kogerentnost' izlucheniya sgustkov (Influence of probability density of electrons on coherence of the bunch radiation),'' Zh. Tech. Fiz. \textbf{47}, 1583--1597 (1977) [Sov. Phys. Tech. Phys. \textbf{22}, 917 (1977)].

\bibitem{Shibata1992} Y.~Shibata, K.~Ishi, T.~Takahashi, T.~Kanai, M.~Ikezawa, K.~Takami, T.~Matsuyama, K.~Kobayashi, and Y.~Fujita, ``Observation of coherent transition radiation at millimeter and submillimeter wavelengths,'' \href{https://doi.org/10.1103/PhysRevA.45.R8340}{Phys. Rev. A \textbf{45}, R8340 (1992)}.

\bibitem{Gevorgyan1982} L.~A.~Gevorgyan and N.~K.~Zhevago, ``Kogerentnoe izluchenie ėlektronnykh sgustkov v lazerakh na svobodnykh ėlektronakh (Coherent radiation of electron clusters in free-electron lasers),'' \href{http://www.mathnet.ru/links/3bab5542d1d54c320b9982a9f39d6239/dan45760.pdf}{Dokl. Akad. Nauk SSSR \textbf{267}, 599--601 (1982)}.

\bibitem{Mkrtchyan1998} 
A.~R.~Mkrtchyan, L.~A.~Gevorgian, L.~S.~Grigorian, B.~V.~Khachatryan, and A.~A.~Saharian, 
``Coherent diffraction radiation from an electron bunch.''
\href{https://doi.org/10.1016/s0168-583x(98)00254-7}{Nucl. Instr. and Meth. B \textbf{145}, 67 (1998)}.

\bibitem{Saharian2001} 
A.~A.~Saharian, A.~R.~Mkrtchyan, L.~A.~Gevorgian, L.~S.~Grigoryan, and B.~V.~Khachatryan,  
``Radiation from an electron bunch flying over a surface wave,''
\href{https://doi.org/10.1016/s0168-583x(00)00288-3}{Nucl. Instr. and Meth. B \textbf{173}, 211 (2001)}.

\bibitem{Gevorgian1993} L.~A.~Gevorgian and A.~H.~Shamamian, ``Coherent amplification of radiation of electron beams interacting with laser beat waves,'' Intern. J. Mod. Phys. A \textbf{28}, 1175 (1993).

\bibitem{Shamamian2007} A.~Shamamian and L.~Gevorgian, ``Coherent sub-millimeter undulator radiation from modulated electron beam,'' International Conference on Charged and Neutral Particles Channeling Phenomena II, \href{https://doi.org/10.1117/12.742068}{Proc. SPIE \textbf{6634}, 66341C (2007)}.

\bibitem{Rosenzweig1995} 
J.~Rosenzweig, G.~Travish, and A.~Tremaine, ``Coherent transition radiation diagnosis of electron beam microbunching,'' \href{https://doi.org/10.1016/0168-9002(95)00484-X}{Nucl. Instr. and Meth. A \textbf{365}, 255 (1995)}. 

\bibitem{HGevorgyan2017}
H.~L.~Gevorgyan and L.~A.~Gevorgian, 
``Coherent radiation characteristics of modulated electron bunch formed in stack of two plates,''
\href{https://doi.org/10.1016/j.nimb.2017.02.079}{Nucl. Instr. and Meth. B \textbf{402}, 126 (2017)}.


\bibitem{Gevorgian2013} 
L.~A.~Gevorgian, K.~A.~Ispirian, and A.~H.~Shamamian, ``Crystalline undulator radiation of microbunched beams taking into account the medium polarization,'' \href{https://doi.org/10.1016/j.nimb.2013.02.034}{Nucl. Instr. and Meth. B \textbf{309}, 63 (2013)}. 

\bibitem{Ispirian2013} 
K.~A.~Ispirian, ``Coherent X-ray radiation produced by microbunched beams in amorphous and crystalline radiators,'' \href{https://doi.org/10.1016/j.nimb.2013.01.072}{Nucl. Instr. and Meth. B \textbf{309}, 4 (2013)}. 

\bibitem{Gevorgian2005}
L.~A.~Gevorgian, ``High acceleration rates obtained in plasma-loaded FEL,'' International Conference on Charged and Neutral Particles Channeling Phenomena, \href{https://doi.org/10.1117/12.640014}{Proc. of SPIE \textbf{5974}, 59740V-1 (2005)}.

\bibitem{Kostyuk2013}
A.~Kostyuk, A.~Korol, A.~Solov’yov, and W.~Greiner, ``Crystalline Undulator: Current Status and Perspectives'' in \emph{Exciting Interdisciplinary Physics. Quarks and Gluons / Atomic Nuclei / Relativity and Cosmology / Biological Systems}, ed. by W. Greiner, \href{https://doi.org/10.1007/978-3-319-00047-3_32}{pp. 399--409 (2013)}.

\bibitem{Kostyuk2011}
A.~Kostyuk, A.~V.~Korol, A.~V.~Solov’yov, and W.~Greiner, ``Demodulation of a positron beam in a bent crystal channel,'' \href{https://doi.org/10.1016/j.nimb.2011.04.006}{Nucl. Instr. Meth. B \textbf{269}, 1482 (2011)}. 

\bibitem{Kostyuk2010}
A.~Kostyuk, A.~V.~Korol, A.~V.~Solov’yov, and W.~Greiner, ``Stable propagation of a modulated positron beam in a bent crystal channel,'' \href{https://doi.org/10.1088/0953-4075/43/15/151001}{J. Phys. B \textbf{43}, 151001 (2010)}. 

\bibitem{Biryukov2007}
V.~M.~Biryukov, ``Comment on ``Feasibility of an electron-based crystalline undulator'','' \href{https://doi.org/10.48550/arXiv.0712.3904}{arXiv:0712.3904 [physics.acc-ph] (2007)}.

\bibitem{Bellucci2003}
S.~Bellucci, S.~Bini, V.~M.~Biryukov, Yu.~A.~Chesnokov, S.~Dabagov, G.~Giannini, V.~Guidi, Yu.~M.~Ivanov, V.~I.~Kotov, V.~A.~Maisheev, C.~Malagù, G.~Martinelli, A.~A.~Petrunin, V.~V.~Skorobogatov, M.~Stefancich, and D.~Vincenzi, ``Experimental study for the feasibility of a crystalline undulator,'' \href{https://doi.org/10.1103/PhysRevLett.90.034801}{Phys. Rev. Lett. \textbf{90}, 034801 (2003)}.



\bibitem{Jackson1999}
J.~D.~Jackson, \emph{Classical Electrodynamics (3rd Ed.)} (John Wiley \& Sons, 1999).






\bibitem{Bazilev1987} 
V.~A.~Bazilev and N.~K.~Zhevago, \emph{Izluchenie Bystrykh Chastits v Veshchestve i vo Vneshnikh Polyakh (The Radiation of Fast Particles in Medium and in External Fields)} (Nauka, Moscow, 1987).

\bibitem{avakian2013}
R.~Avakian, L.~Gevorgian, and L.~Hovsepyan, ``Comparison of the SLAC experimental data on the radiation of planar channeled positrons with theory taking into account the medium polarization,'' \href{https://doi.org/10.1016/j.nimb.2013.02.033}{Nucl. Instr. and Meth. B \textbf{309}, 20 (2013)}.

\bibitem{HGevorgyan2021a}
H.~L.~Gevorgyan, ``X-ray crystalline undulator radiation in water window,'' \href{https://doi.org/10.52853/18291171-2021.14.2-105}{Arm. J. Phys. \textbf{14}, 105 (2021)}.

\bibitem{HGevorgyan2021b}
H.~L.~Gevorgyan, L.~A.~Gevorgian, and A.~H.~Shamamian, ``Gain and features of radiation of positrons in a crystalline undulator with sections,'' \href{https://doi.org/10.3103/S1068337221030117}{J. Contemp. Phys. \textbf{56}, 159 (2021)}.

\bibitem{HGevorgyan2015}
K.~Gevorgyan, L.~Gevorgian, and H.~Gevorgyan, ``Positron bunch radiation in the system of tightly-packed nanotubes,'' \href{https://doi.org/10.48550/arXiv.1512.08282}{arXiv:1512.08282 [physics.acc-ph] (2015)}.

\bibitem{HGevorgyan2024a}
H.~L.~Gevorgyan, K.~L.~Gevorgyan, A.~H.~Shamamian, and L.~A.~Gevorgian, `` Line shape of soft photon radiation generated at zero angle in an undulator with a dispersive medium,'' \href{https://doi.org/10.1016/j.nima.2025.170313}{Nucl. Instr. Meth. A \textbf{1075}, 170313 (2025)}.





\bibitem{Gevorgian2007}
L.~A.~Gevorgian and L.~Hovsepyan, ``Novel effects stipulated by medium polarization at planar channeling positron bunch radiation,'' \href{https://doi.org/10.1117/12.741841}{Proc. SPIE \textbf{6634}, 663408 (2007)}.

\bibitem{Marangos2020}
J.~P.~Marangos, ``Accessing the quantum spatial and temporal scales with XFELs,'' \href{https://doi.org/10.1038/s42254-020-0183-7}{Nature Rev. Phys. \textbf{2}, 332 (2020)}.

\bibitem{Boutet2018}
S.~Boutet and M. Yabashi, ``X-ray free electron lasers and their applications,'' In: S.~Boutet, P.~Fromme, and M.~Hunter, (eds) \emph{\href{https://doi.org/10.1007/978-3-030-00551-1_1}{X-ray Free Electron Lasers}}, pp. 1--21 (Springer, Cham, 2018).

\bibitem{Quiney2010}
H.~M.~Quiney and K.~A.~Nugent, ``Biomolecular imaging and electronic damage using X-ray free-electron lasers,'' \href{https://doi.org/10.1038/nphys1859}{Nature Phys. \textbf{7}, 142 (2010)}.  

\bibitem{Krupyanskii2014}
Y.~F.~Krupyanskii \etal, ``Femtosecond X-ray free-electron lasers: A new tool for studying nanocrystals and single macromolecules,'' \href{https://doi.org/10.1134/s1990793114040046}{Russ. J. Phys. Chem. B \textbf{8}, 445 (2014)}. 

\bibitem{Huang2021}
N.~Huang, H.~Deng, B.~Liu, D.~Wang, and Z.~Zhao, ``Features and futures of X-ray free-electron lasers,'' \href{https://doi.org/10.1016/j.xinn.2021.100097}{The Innovation \textbf{2}, 100097 (2021)}. 

\bibitem{Vartanyants2015}
I.~Vartanyants and O.~Yefanov, ``\href{https://doi.org/10.1201/b15674-13}{Coherent X-ray diffraction imaging of nanostructures},'' In: O.~H.~Seeck and B.~Murphy, (eds) \emph{\href{https://doi.org/10.1201/b15674}{X-Ray Diffraction: Modern Experimental Techniques}}, pp. 341--384 (Taylor \& Francis Group, New York, 2014). 

\bibitem{Yabashi2013}
M.~Yabashi, H.~Tanaka, T.~Tanaka, H.~Tomizawa, T.~Togashi, M.~Nagasono, T.~Ishikawa, J.~R.~Harries, Y.~Hikosaka, A.~Hishikawa, K.~Nagaya, N.~Saito, E.~Shigemasa, K.~Yamanouchi, and K.~Ueda, ``Compact XFEL and AMO sciences: SACLA and SCSS,'' \href{https://doi.org/10.1088/0953-4075/46/16/164001}{J. Phys. B \textbf{46}, 164001 (2013)}. 

\bibitem{Pellegrini2016}
C.~Pellegrini, A.~Marinelli, and S.~Reiche, ``The physics of x-ray free-electron lasers,'' \href{https://doi.org/10.1103/revmodphys.88.015006}{Rev. Mod. Phys. \textbf{88}, 015006 (2016)}. 
























\end{thebibliography}


\end{document}